\documentclass[sigconf,9pt]{acmart}

\AtBeginDocument{%
  \providecommand\BibTeX{{%
    \normalfont B\kern-0.5em{\scshape i\kern-0.25em b}\kern-0.8em\TeX}}}

\setcopyright{rightsretained}
\copyrightyear{2022}
\acmYear{2022}
\acmDOI{10.21428/bf6fb269.58c3a89d}
\acmConference[LIMITS '22]{LIMITS '22: Workshop on Computing within %
  Limits}{June 21--22, 2022}{}
\acmBooktitle{LIMITS '22: Workshop on Computing within %
  Limits, June 21--22, 2022}
\acmISBN{}

\usepackage{verbatim}
\usepackage[textsize=footnotesize, textwidth=15mm]{todonotes}
\presetkeys{todonotes}{fancyline}{}
\usepackage[english=american,babel=false]{csquotes}
\usepackage{nth}

\usepackage{etoolbox}
\newbool{print}
\booltrue{print}

\usepackage{acronym}

\acrodef{DMA}{Direct Memory Access}
\acrodef{ICT}{Information and Communications Technology}
\acrodef{MAC}{Message Authentication Code}
\acrodef{MMIO}{Memory-Mapped I/O}
\acrodef{PCBAC}{Program Counter Based Access Control}
\acrodef{PM}{Protected Module}
\acrodef{PMA}{Protected Module Architecture}
\acrodef{TCB}{Trusted Computing Base}
\acrodef{TEE}{Trusted Execution Environment}
\acrodef{TXT}{Trusted eXecution Technology}
\acrodef{WSN}{Wireless Sensor Network}

\begin{document}

\title{Sustaining Security and Safety in ICT: \\
  A Quest for Terminology, Objectives, and Limits}

\author{Jan Tobias M\"uhlberg}
\email{jantobias.muehlberg@cs.kuleuven.be}
\orcid{0000-0001-5035-0576}
\affiliation{%
  \institution{imec-DistriNet, KU Leuven}
  \city{Leuven}
  \country{Belgium}
  \postcode{3001}
}

\begin{abstract}

Security and safety of system are important and intertwined concepts in the
world of computing.
\notbool{print}{ Where security aims to protect systems from intentional
attacks, safety strives to ensure systems' operation under unintentional
hazards.  With the increasing societal dependence on digital technologies
and the growing threat that changing environmental conditions and cyber
attacks pose to our infrastructures, we understand that safety and security
are linked concepts that critical systems but also consumer electronics
must implement.

} %
In recent years, the terms \enquote{sustainable security} and
\enquote{sustainable safety} came into fashion and are being used referring
to a variety of systems properties ranging from efficiency to
profitability, and sometimes meaning that a product or service is good for
people and planet. This leads to confusing perceptions of products where
customers might expect a sustainable product to be developed without child
labour, while the producer uses the term to signify that their new product
uses marginally less power than the previous generation of that products.
Even in research on sustainably safe and secure ICT, these different
notions of terminology are prevalent.  As researchers we often work towards
optimising our subject of study towards one specific sustainability metric
-- let's say energy consumption -- while being blissfully unaware of, e.g.,
social impacts, life-cycle impacts, or rebound effects of such
optimisations.

In this paper I dissect the idea of sustainable safety and security,
starting from the questions of what we want to sustain, and for whom we
want to sustain it. I believe that a general "people and planet" answer is
inadequate here because this form of sustainability cannot be the property
of a single industry sector but must be addressed by society as a whole.
However, with sufficient understanding of life-cycle impacts we may very
well be able to devise research and development efforts, and inform
decision making processes towards the use of integrated safety and security
solutions that help us to address societal challenges in the context of the
climate and ecological crises, and that are aligned with concepts such as
intersectionality and climate justice. Of course, these solutions can only
be effective if they are embedded in societal and economic change towards
more frugal uses of data and ICT.

\end{abstract}


\maketitle
\renewcommand{\shortauthors}{Jan Tobias M\"uhlberg}


\section{Introduction}

Our total reliance on computing infrastructures becomes most apparent when
%
%
those infrastructures fail. I am writing these lines while waiting at a
train station in The Netherlands, on Sunday the \nth{3} of April 2022.
There is free coffee, but pretty much no trains are operational. \emph{NS},
the main network operator, announced earlier that due to a technical
problem there will be no trains running until 17:00. No further details are
revealed at this moment but speculations are going wild: \enquote{It's
probably an attack,} someone says. Across the country, thousands of people
are stranded at stations. Alternative means of public transport cannot cope
with the rush demand, roads are jammed with taxis and personal vehicles.
\enquote{Due to the enormous impact of the failure in the IT system, it is
unfortunately not possible to run any trains today,} we hear later. The
failure \enquote{affected the system that generates up-to-date schedules
for trains and staff. This system is important for safe and scheduled
operations: if there is an incident somewhere, the system adjusts itself
accordingly. This was not possible due to the failure.}\footnote{
\enquote{Sunday 3 April no more NS trains}, 2022-04-03 19:07 CEST.
Archived at \\
\url{https://web.archive.org/web/20220404172602/https://www.ns.nl/en/travel-information/calamities/sunday-3-april-no-more-ns-trains.html}
%
} At least we had free coffee.

As our society is becoming increasingly dependent on digital technologies,
research and industry made important efforts to ensure the safety and
security of these technologies. Here \emph{safety} is the property of a
system to achieve an acceptable level of risk by controlling recognised
hazards -- predominantly environmental conditions or human error, which are
expected to behave within certain parameters. In the context of \ac{ICT},
safety thus involves properties of the hardware (e.g., ensuring safe
operating temperatures for a processor) and software (e.g., ensure correct
execution and safe behaviour of the software scheduling NS trains and
staff). \emph{Security} is distinguished from safety and denotes the
property of a system to withstand intentional attacks from an intelligent
adversary. Such an adversary will seek to interact with the system in
unexpected ways to make the system misbehave, often violating safety as a
result.

The fundamental distinction between security and safety is regarding the
presence (or absence) of an intelligent
and purposefully malicious adversary, their ability to act outside of the
limits of safe interactions considered when the system was designed. So,
when the Ukrainian power grid was hacked in 2015, the attackers
\enquote{were skilled and stealthy strategists who carefully planned their
assault over many months, first doing reconnaissance to study the networks
and siphon operator credentials, then launching a synchronised assault in a
well-choreographed
dance,}\footnote{\enquote{Inside the Cunning, Unprecedented Hack of
Ukraine's Power Grid}, 2016-03-03. \url{https://www.wired.com/2016/03/inside-cunning-unprecedented-hack-ukraines-power-grid/}}
which left 230,000 residents without electricity.

\subsection{Security and Safety as Emergent Properties}

With this background it becomes clear that security and safety are emergent
properties that only apply to a system that is being used in a specific
context and environment: A processor may safely operate under the expected
outdoor temperatures in Belgium but not in Ethiopia. A control software may
be acceptably secure to operate on the isolated local network of a factory
but not over a public network where an attacker may modify or replay
communication. It also becomes clear that safety and security are heavily
dependent on assumptions that may be violated once in a while, e.g., by
increasingly high summer temperatures in Belgium, or by an attacker who
unexpectedly gains access to a seemingly isolated network. As such, neither
safety nor security are absolute but are about reducing risk levels to be
acceptable by society and within the margins of standards and regulations.

Furthermore we see that safety and security are interconnected, in
particular, that safe operation of equipment can only be guaranteed if that
equipment is also secure. The field of study that combines these two
aspects of risk assessment and engineering is \emph{dependability}
engineering~\cite{national2007dependability}, and important progress has
been made regarding safety and security co-assurance for critical
systems~\cite{gleirscher2022challenges}. Particularly hard to assess in
this context is the long-term survivability of systems, as Eder-Neuhauser
et al. point out with respect to the security of power grid infrastructure
in~\cite{eder-neuhauser_cyber_2017}: \enquote{However, unlike consumer
electronics, traditional power grid environments have a focus on long-term
stability and plan for hardware life-spans of 10 years or more. As devices
age, unknown vulnerabilities of hardware, operating system, software, and
protocols emerge. Such vulnerabilities pose a serious threat to the
infrastructure. While consumer electronics need not fulfil the same
life-cycle requirements of industrial devices, their base technology is
similar.}

\subsection{This Paper}

In this paper I highlight and discuss the relevance of security and safety
for sustainability in the ICT sector. Recently, more researchers and
companies strive to link their output -- be it research, equipment or
services -- to varying definitions of sustainability, based on
characteristics and measurements that involve efficiency, overheads,
profitability, or environmental impacts. Reflecting on these
characteristics, I analyse the idea of sustainable safety and security,
starting from the questions of what we want to sustain, and for whom we
want to sustain it. I hope that this discussion can contribute towards
developing an integrated understanding of safety and security that helps us
to address sustainability topics and societal challenges in the context of
the climate and ecological crises. I believe that this understanding may
then allow us to align technical development with concepts such as
intersectionality and climate justice. What becomes clear is that
sustainability efforts that target security- and safety engineering can
only be effective if they are embedded in societal and economic change
towards more frugal uses of data and ICT.

\section{Notions of Sustainability in the Context of Security and Safety}

In~\cite{pavert_sustainable_2019}, Paverd et al. link security and safety,
and implicitly dependability, to a notion of \emph{sustainability} that
focuses on maintaining key properties of systems by preparing them for
threats not known today, defining design principles for a vision of
\emph{sustainable security and safety}. These design principles focus on
diversification, replicability and relocatability of resources, containment
of failures, adaptability and updatability, with simplicity, verifiability
and minimisation of assumptions as overarching concepts. While there are
many open research challenges regarding the implementation of these
principles, following the vision of Paverd et al. would lead to the
development of software and equipment designed for safe and secure
longevity.

\subsection{On Costs and Benefits}

System designs resulting from following the sustainable security and safety
principles from Paverd et al.~\cite{pavert_sustainable_2019} would likely
incur increased costs regarding the development, initial equipment
purchase, and the maintenance of installations. For example, modern
processors typically come with built-in support for encryption. Since about
2010, processor vendors implement instructions for the AES
cipher~\cite{daemen1999aes} in hardware, and ever since viable attacks
against these hardware implementations are being published. These attacks
necessitate the replacement of equipment where the risk of such attacks is
deemed unacceptable. A way to potentially limit hardware replacement is by
using updatable hardware and crypto agility, which may come with
different security guarantees and more complex equipment in the first
place.

Another well-known example for such issues are the Spectre and
Meltdown vulnerabilities~\cite{kocher_2018spectre,lipp_2018meltdown}, which
render entire generations of processors vulnerable to software-level
attacks.  Processor vendors and operating system developers have provided
patches to mitigate these vulnerabilities at the expense of a substantial
degradation of performance and increase of energy
consumption~\cite{herzog_price_2021,prout_measuring_2018,alhubaiti_impact_2019}.
With application-specific energy overheads of up to
72\%~\cite{herzog_price_2021}, applying these patches in existing
installations raises questions of resource availability and scalability,
and may necessitate a lot of additional equipment to be added to compensate for the
loss of processing power. Indeed, systems that are designed by following
the sustainable security and safety principles
from~\cite{pavert_sustainable_2019}, in particular diversifiability,
relocatability, and containment of breaches, can very likely be adapted and
kept secure under this changing attacker model, but this maintenance will
still incur substantial costs.

We realise that, with layers and layers of
patches and security features built on top of each other,
and that all require continuous maintenance and re-evaluation to maintain
reasonable levels of security for internet-connected systems, these costs
build up quickly and contribute to the overall short lifespan of ICT
equipment. 

\subsection{Sustainability as a Requirement}

Sustainability as a first-class quality of \ac{ICT}, on an equal footing
with safety and security, has first been proposed by Penzenstadler et al.
in~\cite{penzenstadler_safety_2014}, arguing that \enquote{instead of
merely optimising current systems, software engineers must embrace
transition engineering -- an emerging discipline that enables change from
existing unsustainable systems to more sustainable ones by adapting and
filtering demand to a declining supply.} Here, transition engineering is a
discipline that aims at \enquote{identifying unsustainable aspects of
current systems, assessing the risks posed by those aspects, and
researching and developing ways to mitigate and prevent systemic failures
through adaptations}~\cite{krumdieck_survival_2011}. Penzenstadler et al.
point to a conflict between organisational goals on the one side and
security, safety, and sustainability on the other, because these qualities
are commonly perceived as barriers to profitability; predominantly so
because companies do not pay for environmental and social impacts of their
products. In this context, sustainability can be seen as a non-functional
requirement similar to security or safety, the implementation of which
requires the establishment of a sustainability culture in organisations and
in society.

There is a growing body of literature on requirements engineering for
%
%
software systems that engage with the multi-dimensional nature of
sustainability, e.g.~\cite{becker_requirements_2016,venters_software_2018,duboc_we_2019}.
Albeit with no specific considerations for safety and security, it is noted
that systems resulting from such efforts are \enquote{different when
sustainability principles and therefore long-term consequences are
considered}~\cite{becker_requirements_2016}. As safety and security are
emergent properties of systems that involve qualities of, and interactions
between, a system's hardware, a system's software, and the operating
environment, engineering processes that aim at sustainably safe and secure
systems need to involve approaches to hardware-and-software co-design based
on requirements catalogues such as Paverd et
al.~\cite{pavert_sustainable_2019}. Notions of sustainability and
circularity in micro electronics that consider the physical basis of ICT
infrastructures have, e.g., been developed by Griese et al.
in~\cite{griese_sustainable_2004}, and Clemm et al.
in~\cite{hu_implications_2019}, which need to be integrated with the work
on sustainability for software systems mentioned above.

Earlier notions of sustainability in ICT lead back to K\"ohler and
Erdmann~\cite{kohler_expected_2004} and Hilty et
al.~\cite{hilty_relevance_2006}, who distinguish three dimensions where
software systems impact sustainability: \emph{(1)} effects of the physical
existence of ICT, such as the impacts of the production, use, recycling and
disposal of equipment; \emph{(2)} indirect environmental effects of due to
changed processes in other sectors, including e.g., the optimisation of
industrial processes or uses of products and services which in turn result
in changed environmental impacts of these processes, products or services;
and \emph{(3)} tertiary effects resulting from the medium- or long-term
adaptation of behaviour such as consumption patterns or business models, as
a result of the continuous availability of ICT products and services. In
summary, Hilty et al.~\cite{hilty_relevance_2006} conclude that
\enquote{the overall impact of ICT on most environmental indicators seems
to be weak, the impact of specific areas or types of ICT application can be
very relevant in either direction. On an aggregated level, positive and
negative impacts tend to cancel each other out.} More research is certainly
necessary to understand whether these conclusions still hold in the light
of current socio-economic and environmental impacts of ICT sector, for
example the online advertising
sector~\cite{frick_online_2021,cucchietti_carbolytics_2022}, being an
engine of economic growth and a key protagonist to sufficiency-oriented
consumption.

The above definitions and aspects of sustainability are all focused on
\acp{ICT}. The most generally applicable, and certainly the most quoted
notion of \emph{sustainable development} (not sustainability in general)
comes from Brundtland~\cite{brundtland_our_1987}, referring to
\enquote{development that meets the needs of the present without
compromising the ability of future generations to meet their own needs.} In
difference to the definitions discussed before, Brundtland puts an emphasis
on societal needs. If these are no longer satisfiable, what we understand
as society right now, may cease to exit. Thus, maybe we should \emph{not}
see sustainability as a quality of ICT systems, besides security and
safety, but use Brundtland's definition as the overarching safety
requirement for all ICT development?

\subsection{Conflicting Expectations}

While all these diverging notions of sustainability and sustainable safety and security make
sense with respect to their specific frames of reference, they are surely
confusing for everyone who simply wants to buy a sustainable product or use a
sustainable service. As suggested by sparsely available empirical research,
public understanding of sustainable development may centre around
\enquote{green issues,} with a focus on saving energy and reducing waste
(cf. \cite{scott2015public}). Of course, other interpretations are possible
and one might expect a product to be manufactured under fair labour
conditions or with responsibly mined resources. Now, while a manufacturer
might label their business as sustainable on the basis that they are able
to continuously increase employee salaries, a consumer might expect the
product to be somehow \enquote{good for the planet.} To make things worse,
anecdotal evidence shows that there is also \enquote{secret
sustainability,} where companies do not announce their sustainability
efforts out of fear that consumers would expect their more sustainable
products to be inferior to less sustainable competitors, either in terms of
a reduction in product quality, or an increase in the price of
manufacturing.\footnote{\enquote{Why industry is going green on the quiet},
2019-09-08. \url{https://www.theguardian.com/science/2019/sep/08/producers-keep-sustainable-practices-secret}}

Of course, all these notions and considerations are fairly generic. What
could it now possibly mean if a product or service features sustainable
security and safety? Does it mean that the products jointly manages risks
related to safety, security, and sustainability, that it sustains its
safety and security properties, or that it implements safety and security
in some way that is good for \enquote{people and planet?}
I want to dissect
this question by asking explicitly what it is that we sustain, and for whom
we sustain it.\footnote{Tainter~\cite{tainter_framework_2003} was likely
the first to ask these questions to reason about sustainability efforts:
\enquote{Directing sustainability efforts in productive directions, then,
requires understanding that it is a matter of values, not invariant
biophysical processes. Some people and some ecosystems benefit from
sustainability efforts, while others don’t. When confronted with the term
'sustainability,' therefore, one should always ask: Sustain what, for whom,
for how long, and at what cost?} Benessia and
Funtowicz~\cite{benessia_sustainability_2015} use this framework to analyse
\enquote{ways in which techno-science modifies and determines the object
and the subject of sustainability.}}
Understanding that personal safety and security, together with privacy, are
fundamental human right listed in the Universal Declaration of Human Rights
(UDHR) makes asking these these questions even more important: We cannot
barter notions of sustainability for safety and security. Instead we must
strive to design and use technologies so as to reduce the divide between
the over-served and those who are marginalised by society, economic
development, or technology.

\section{Sustaining What? And for Whom?}

Predicting the impact of security and safety  technologies on the different
dimensions of sustainability -- society, environment, culture, and economy
-- is difficult. Thus, it is not surprising that research in security and
safety hardly ever bothers with such analyses but focuses on reporting more
easily quantifiable evaluation metrics such as security under different
attacker models, performance and overheads, system complexity or the price
of installations.

But would an analysis beyond these metrics even be useful? Is it not the
case that managing security and safety risks in any product will always
incur overheads, will always increase the environmental footprint of a
product, and will always ask people to change their behaviour? Does
\emph{sustainable security and safety} then make sense as a term or
concept, or is it merely a framing that we as security and safety
professionals use to align our work with a current marketing trend that
aims to make consumers feel good with their sustained consumption? Or should
we seek to carbon-offset the impacts of security and safety mitigations and
report in our papers the number of trees we planted for every
microcontroller purchased for some experiment?

I fundamentally believe that sustainability in the general sense of
\enquote{meeting the needs of the present without compromising the ability
of future generations} is not a property of an individual engineering
discipline or industry sector. Even if the ICT sector would reach net-zero
emissions, it would still not be sustainable in itself due to its demand
for virgin resources and exploitative labour, and the environmental and
social impacts related to these.%
\footnote{The IPCC WG3 AR6 report suggests that net-zero industries are
challenging but feasible is feasible, specifically regarding greenhouse gas
emissions, requiring coordinated action: \enquote{Net-zero $CO_2$ emissions
from the industrial sector are challenging but possible. Reducing industry
emissions will entail coordinated action throughout value chains to promote
all mitigation options, including demand management, energy and materials
efficiency, circular material flows, as well as abatement technologies and
transformational changes in production processes.}~\cite{grubb2022climate}}
We understand that as a society we can reach and maintain Brundtland's
notion of sustainable development by operating within the planetary
boundaries~\cite{rockstrom_planetary_2009}, by defining our economic
activities in the doughnut~\cite{raworth_doughnut_2017}, and we understand that \acp{ICT}
can contribute to this: \enquote{Infrastructure design and access, and
technology access and adoption, including information and communication
technologies, influence patterns of demand and ways of providing services,
such as mobility, shelter, water, sanitation, and nutrition. Illustrative
global low demand scenarios, accounting for regional differences, indicate
that more efficient end-use energy conversion can improve services while
reducing the need for upstream energy by 45\% by 2050 compared to
2020.}~\cite{grubb2022climate}

The notion of sustainability that
relates most closely to ICT systems follows
Tainter~\cite{tainter_social_2006}, who sees \enquote{sustainability is an
active condition of problem solving, not a passive consequence of consuming
less.} Ultimately it is the responsibility of society and our democratic
processes to define what resources we are willing to spend on critical
infrastructures and for individual consumption, and which role security and
safety needs to play in this framework. And if we as society, in such a
modelling and planing exercise, and with the necessary democratic
legitimacy, conclude that every citizen may procure a new smartphone every
ten years (following a model from~\cite{millward-hopkins_providing_2020}),
then it must be the task of the ICT sector to design a device that can be
operated with adequate security guarantees -- defining that term also
requires democratic legitimacy -- for at least the anticipated duration of
use.

However, the reach of ICT is pervasive and few industries have shaped
society as much as modern communication systems and automation. This huge
societal impact must come with equally huge responsibility, and as researchers,
engineers, or companies active in this field, we occupy a position of power
that we must use and strive to provide adequate notions of safety and
security that are inclusive, that serve society as a whole, and that do not
make parts of the world pay for the advancement of others.

\subsection{Engineering Practice: Cats and Mice}

The filed of systems security is particularly remarkable in this regard:
Security is generally perceived as a cat-and-mouse game where whatever
security measures are being put in place and are (almost always) defeated
by a sufficiently advanced and resourceful attacker in the foreseeable
future. As such, it is impossible to infinitely sustain the security of
some system, nor is is possible to make security sustainable in the
\enquote{people-and-planet} sense.  In~\cite{muehlber_2018post_meltdown} we
argue, with respect to processor security in the context of the
aforementioned Spectre and Meltdown attacks, that this cat-and-mouse cycle
can possibly slowed down by following a more principled approach to
designing security solutions, which would likely follow the criteria for
\enquote{sustainable security and safety} laid out by Paverd et
al.~\cite{pavert_sustainable_2019}.

Nevertheless, sustaining security would still be energy- and material
intensive. We have to understand that, on the one hand, security mechanisms
may serve a very specific demographic, namely those who can afford to care
and for whom the economic benefit of expensive security comes with the
promise of even greater return on investment. On the other hand,
implementing security that is designed for this one specific demographic,
may put a potentially much larger group of people at the risk of extended
exploitation: for example those people whose digital security is not deemed
worthy of the investment, those who work in artisanal mines in the global
south, extracting the raw materials for new and more secure electronics
without ever being able to afford a device featuring the latest security
features, and also those for whom a new security solution may require
unreasonable adaptation of their social behaviour. As Wu et al. point out
in~\cite{wu_sok_2022}, \enquote{ignoring human social behaviours in
designing [security and privacy] systems leads to maladaptive user
behaviours that either reduce security, cause social friction, or both.}

Interestingly, safety emerged as a very different discipline of
engineering.  Being focused on mitigating adverse environmental conditions
and human error within constrained parameters -- users make benign errors
but do not actively seek to disrupt the functioning of a system -- safety
engineering did traditionally not exhibit the characteristics of the
aforementioned cat-and-mouse game as much as security does. As such,
managing safe operation of a system  requires less of a continuous effort:
Once the safety requirements for a specific operating environment and risk
vectors are established and implemented, the system will only require
maintenance when components fail. Under harsh conditions that may of course
happen regularly, but it is still predictable and the mitigations to
failure -- maintenance windows, replacement components, etc. -- can be
planned in advance. This approach to safety, however is changing. With the
increased connectivity and distributedness of safety-critical systems,
these systems must focus more and on dependability and integrate security
engineering as a prerequisite for system safety.

\subsection{The \enquote{What}}

Therefore the key to \emph{sustaining security and safety} probably is to
be very cautious about defining the assets that need protection: \emph{What
do we want to sustain?} The business models of companies? The security of a
smart grid infrastructure that enables a community to optimally use
renewable energy sources? The confidentiality of amassed profiling data and
machine learning models, collected and processed with the purpose of
learning the data subjects' intimate desires to produce targeted
advertising and sell them more things to perpetuate economic growth? Threat
modelling~\cite{shostack2014threat} is the discipline of security
engineering that is meant to give us an understanding of assets, actors,
and the required threat mitigations, albeit without a critical reflection
on whether these assets need to be there in the first place. What is
required on top of this is a notion of critical refusal, a challenging
reflection on harmful practices of building infrastructure, data collection
and the like, while opening up spaces to develop alternative proposals (cf.
\cite{simpson2007ethnographic, theilen_feminist_2021}). Such notions of
refusal have, e.g., been expressed in the Feminist Data
Manifest-No~\cite{cifor2019feminist}, which could very well be appropriated
by researchers and practitioners in safety and security engineering.

Importantly, alternative approaches in this domain do not exclude new ICT
systems or the data-based optimisation of businesses or data per se. Moving
towards data collections and inference that serves the public interest,
that is being collected and processed following data protection and privacy
by-design principles~\cite{gurses2011engineering}, and that is then being
made available under an open data policy, could already lead to enormous
reductions in the need for security infrastructure. That is because such an
approach would enable sharing and reuse of data and infrastructures, and
would reduce the need for confidentiality protection, with only the need
for integrity protection and availability remaining.

Reflections towards critical refusal of technical developments and towards
reducing the quantity of assets that need protection, essentially enabling
us to, e.g., switch data centres off instead of patching thousands of
machines and then mitigating the overheads by adding more machines,
potentially have a tremendous positive impact on societies digital security
requirements and also reduce environmental impacts dramatically, much more
than technical optimisations in the implementation of cryptographic
algorithms and the like. Also, if protecting the confidentiality of assets
is no longer a security objective that need to be upheld for a majority of
assets, it may be feasible to operate (parts of) data centres without
patching confidentiality-related vulnerabilities such as, in certain
scenarios, Meltdown and Spectre.

\subsection{The \enquote{Who}}

Technology affects people differently. If we, for example, develop a
product that aims to sustain road safety for fast vehicles by introducing a
vehicular communication system that allows for coordinated emergency
braking~\cite{cao20165g}, we probably impose additional risks on road users
to whom this technology is not available. Pedestrians, cyclists, and users
of older vehicles may see themselves endangered by a new generation of
faster vehicles whose drivers show little concern for the presence of
vulnerable road users.
Or we could think of an implementation of onion
routing~\cite{goldschlag1999onion} to sustain users' privacy when browsing
the Internet, which comes with increasing material and energy consumption
due to additional layers of cryptography and additional routing hops for
network packets. In consequence, this privacy tool may not become equally
available to everyone and may never benefit, or indeed impose additional
burdens, on the most vulnerable members of society, internationally. Yet,
the availability of this tool has been important, e.g., in the context of
whistleblower protection and to help those organising and fighting
oppressive regimes.

Specifically when looking at the North-South divide regarding the
availability of technology we see dramatic differences in who has access to
which technologies, whom technological advancement serves, whom we develop
technology for, and respectively who controls technology roll-out and
platforms. This all affects how we define sustainability in ICT. My
institution, for example, collaborates with an initiative to collect and
refurbish ICT equipment, typically after three of four years of university
use, to give them a second life in the Global
South.\footnote{\enquote{Life cycle management of laptops and desktops},
2022-04-03. \url{https://admin.kuleuven.be/icts/english/sustainability/laptops-desktops}}
For what looks like a great initiative at a first glance, we have to ask if
this practise can serve as a sustainability goal or if it instead indurates
current power structures. Why is it that people in the Global South are
supposed to be fine with four-year-old equipment while we cannot use that
equipment any further? Why is it that those economies, where most of the
mining of basic resources for our equipment is done, at horrendous social
and environmental costs, cannot afford new equipment but depend on our
donations? Can safe and sustainable waste management be guaranteed when the
equipment eventually reaches its end of life in the destination country?
How is security managed for used equipment where hardware and built-in
software or credentials could be compromised?  Do such equipment donations
actually aim at mitigating power imbalances between the Global North and
the Global South?


%

\section{A Concrete Example}

Last year, my colleagues and I published a paper~\cite{alder_2021_aion} in
%
%
which we develop a security architecture that provides a notion of
guaranteed real-time execution for dynamically loaded programs that are
isolated in so-called enclaves. These enclaves are are a concept developed
in the context of trusted computing -- techniques to execute code in
isolation from other software, even from the operating system. On top of
isolation, trusted execution environments provide primitives to convince a
remote party of the integrity of the enclave and whatever computations this
enclave might perform (cf.~\cite{maene:hardware_comparison} for a survey).
We implement preemptive multitasking and restricted atomicity on top of
these enclaves, illustrating separation of concerns where the hardware is
to enforce confidentiality and integrity protections, while a small
enclaved scheduler software can enforce availability and guarantee strict
deadlines of a bounded number of protected applications. The approach is
designed for open systems, where multiple distrusting actors may run
software on the same processor, and without introducing a notion of
priorities amongst these mutually distrusting applications. We implement a
prototype on an extremely light-weight open-source processor, and
illustrate in a case study that protected applications can handle
interrupts and make progress with deterministic activation latencies, even
in the presence of a strong adversary with arbitrary code execution
capabilities. We envision our processor designs to be used in the context
of heterogeneous control networks that implement distributed sensing and
actuation for critical infrastructures, for example in the context of smart
farming or to facilitate the transition towards renewable energy sources.
In~\cite{scopelliti2021hetero} we outline such a scenario.

Our two papers, \cite{alder_2021_aion} and \cite{scopelliti2021hetero},
come without a sustainability evaluation, and I have not seen a conference
in the security, safety, and dependability domain that asks for such an
evaluation to be part of submissions. One reason for this is probably the
lack of established criteria to do so. In our case, evaluating our work
along the lines of the relevant aspects of the sustainable security and
safety principles laid out by Paverd et al.
in~\cite{pavert_sustainable_2019} would be a reasonably thing to do.
Alternatively, one could also look at specific use cases and argue how
these contribute towards achieving the UN's Sustainable Development
Goals~\cite{mata_role_2016}. With funding agencies beginning to ask for an
alignment of research proposals with the Sustainable Development Goals, the
latter might become a common approach that could  very well be picked up by
conferences in the future.

Yet, my idea to ask explicitly what we strive to sustain and for whom we
aim to sustain it goes further than the above approaches, in the sense that
it implies the question of whether our work may incur harms to certain
demographics throughout a system's life cycle. I.e., it is easy to justify
that a system potentially does some good but it requires a structural
framework -- similar safety arguments or threat modelling -- to argue that
a system does no harm.

\subsection{Assessing the Solution}

For our research, I would say that \emph{we strive to sustain the secure
and safe operation of of smart actuation and sensing equipment that aims to
optimise the distribution and use of resources.} We thought specifically of
use cases where very little computational resources are required to perform
the control task, where distributed processing is useful but local control
is needed as a safety mechanism to handle emergencies, and where equipment
should be capable of receiving updates and reconfiguration over decades.
Specifically, we thought of an underground irrigation system in a fruit
orchard at a community farm I am involved with. With increasing water
stress in Belgium, irrigating the orchard may be necessary in the near
future. Yet, doing so with drip irrigation would be wasteful, and frequent
disturbance of the top soil and its vegetation needs to be avoided as it forms
a valuable ecosystem that prevents loss of soil moisture and erosion. With
these requirements in mind, we extended an extremely light-weight processor
-- with a predicted per-unit price of around USD 1.- -- with very strong
security and safety features.

But whom does it serve, for whom do we sustain something? \emph{We envision
our technology to be rolled out as a long-term investment in sectors that
work towards climate mitigations. We therefore seek to minimise complexity
and per-unit price, and open-sourced our designs and prototypes.} Of
course, we still proposed a new hardware design, which needs to be produced
by suppliers that may perpetuate global power imbalances. Yet, with the
open-designs and tooling, we hope that our approach lowers the entrance
barrier for experimenting with the technology or to even build complex
scenarios based on outdated field-programmable hardware (FPGAs) typically
used in prototyping. The concern that remains is that our technology may
not become easily available to rural communities in the global south that
are most vulnerable and most affected to climate change. There is also the
concern that, depending on the application domain, distributed control in
this context unnecessarily exposes infrastructure to network-level attacks
and failures. Our work does, however, put an emphasis on availability in
local control systems, enabling the development of secure control systems
that can be isolated from network-level attackers and that strive to
minimise assets that need protection without preclude larger deployment
scenarios such as sensor networks.

\subsection{Assessing the Use Case}

The potential use case from~\cite{alder_2021_aion}, smart farming, has been
investigated regarding sustainability properties in related work.
In a critical reflection of smart farming, Streed et
al.~\cite{streed_how_2021} conclude that \enquote{agricultural system
planners need to think in a way that holistically addresses all the
services that society desires, not marginally improve a system that fails
to deliver all of these service [\ldots]. Advances in data collection and
analytic techniques provide a valuable opportunity to re-envision
agriculture in ways that have never before been tried, that more closely
mimic nature~\cite{thrall_darwinian_2013}, or that maximise robotics and
technical solutions. There is a need to combine computing and human action
and desires into agriculture, but it should be done wisely, not 'smart'ly.}
These observations are in-line with the latest IPCC WG3 report, reporting
that \enquote{Taken together, [precision agriculture] technologies provide
farmers with a decision-support system in real time for the whole farm.
Arguably, the world could feed the projected rise in population without
radical changes to current agricultural practices if food waste can be
minimised or eliminated. Digital technologies will contribute to
minimising these losses through increased efficiencies in supply chains,
better shipping and transit systems, and improved
refrigeration.}~\cite{grubb2022climate} Following these analysis, our work
aims to provide secure and safe ICT support for socio-technical systems
that involve high agroecological
complexity~\cite{raghavan_computational_2016} such as permaculture.

\subsection{Abuse Cases and Harmful Impacts}

Beyond our best intentions, we cannot avoid the following harmful abuse
cases of our work: \emph{(1)} implementing systems that facilitate
oppression such as sensing and actuation in surveillance, policing, or
migration control; \emph{(2)} hiding unintended code, e.g., malware,
in secured applications; \emph{(3)} abuse of security features as digital
rights management or to impede update, repair, or reuse of systems. These
challenges can only be overcome by regulation and approaches to community
governance of digital infrastructures.

\section{Conclusions}

For years already we find ourselves in a permacrisis of wars, humanitarian
emergencies, and climate and ecological collapse. The IPCC's sixth
assessment report \enquote{Climate Change 2022: Impacts, Adaptation and
Vulnerability} issues a stern warning:
\enquote{The cumulative scientific evidence is unequivocal: Climate change
is a threat to human well-being and planetary health. Any further delay in
concerted anticipatory global action on adaptation and mitigation will miss
a brief and rapidly closing window of opportunity to secure a liveable and
sustainable future for all.}~\cite{poertner2022climate}
In this context, I feel that a strong notion of environmental
sustainability should form the baseline of a system's safety requirements.
That is, if \enquote{concerted anticipatory global action on adaptation and
mitigation} to the climate and ecological emergency is not an inherent goal
of the system, then using the system does involve unacceptable risks to
human wellbeing, and the system should no longer have a place in our
society. Working towards these goals, I argue that we as researchers,
engineers, and as society as a whole, need to follow a path of critical
refusal, reflecting on digital infrastructures, their importance for human
wellbeing. 

\subsection{Sustainability for Safety and Security}

With respect to \acp{ICT}, and to security and safety in particular,
%
%
sustainability concerns must always be reflected in the context of the
systems and societal function that we aim to implement or move into the
digital world. When designing digital infrastructures we must never
compromise safety and security (and neither privacy) and the equitable and
just implementation of -- and access to -- these (cf. design
justice~\cite{costanza-chock_design_2020}), in favour of narrow notions of
sustainability. However, if we accept sustainability as a multidimensional
\enquote{active condition of problem solving, not a passive consequence of
consuming less}~\cite{tainter_social_2006}, we see a large window of
possible design choices, not only technological ones, opening up. This
leaves us with the problem of complexity regarding system design, system
costs, and the impact of safety and security decisions on the system, its
users, and other stakeholders. Tainter~\cite{tainter_social_2006} hints
towards a number of approaches to deal with this complexity. These range
from \enquote{don't solve the problem} over approaches that \enquote{shift
and defer costs} of complexity, to ideas that \enquote{connect costs and
benefits} and \enquote{revolutionise the activity}. All may be valid, and
-- specifically regarding safety, security, and privacy -- established risk
assessment methodologies allow us to adequately assess complexity and
potential harms for different pathways, and to guide decision making.

As an attempt to give a definition I propose the following: \emph{ICT
systems that sustainably incorporate safety and security are designed to
minimise the risk of hazards for all parties involved in the life cycle of
the system, while maximising the safe and secure life span and
possibilities for reuse of that system and its components.}

\subsection{\ldots and what to do with that definition?}

There are numerous additional terms, technologies and concepts that I could
possibly mention to make my definition more concrete: from design justice
to open-sourcing hardware and software, to recommendations for
implementation and verification strategies. None of these concepts would be
directly applicable.

However, because maintaining (i.e., sustaining) security and safety of a
system will necessarily increase the overall environmental footprint of
that system, we should strive to minimise the number of assets that
actually require protection:%
\footnote{The latest IPCC WG3 report acknowledges this in a more general
context: \enquote{Demand-side mitigation encompasses changes in
infrastructure use, end-use technology adoption, and socio-cultural and
behavioural change.  Demand-side measures and new ways of end-use service
provision can reduce global GHG emissions in end use sectors by 40-70\% by
2050 compared to baseline scenarios, while some regions and socioeconomic
groups require additional energy and resources. Demand side mitigation
response options are consistent with improving basic wellbeing for
all.}~\cite{grubb2022climate}}
More than 80\% of all data ever created, consumed, and stored has been
accumulated in just the last five years.%
\footnote{\url{https://www.statista.com/statistics/871513/worldwide-data-created/}}
As our society becomes obsessed with data collection, i.e. the production
of new digital assets, many of which will eventually be compromised,%
\footnote{Data breaches strongly correlate with with the growing number of
assets:
\url{https://www.informationisbeautiful.net/visualizations/worlds-biggest-data-breaches-hacks/}}
we are missing out on opportunities do build more frugal infrastructures
that can be sustainably secured and that sustainably protect people and
communities.
Therefore, the key to decisions about
what systems we implement and how we regulate these systems must be
democratic processes that reflect on critical societal functions and their
allocated resources within the planetary boundaries.

Importantly, following the above definition and ideas for sustaining safety
%
%
and security does not ignore industrial practice but instead asks for
revised approaches to requirements engineering, engineering practices and
regulatory frameworks and industry
standards that consider safety and security risks beyond immediate hazards
but follow inclusive life-cycle assessments of technologies and systems.

Then, following ideas from research in software
sustainability~\cite{venters_software_2018,duboc_we_2019} and devising
equipment that generally has a low materials and energy footprint, that can
be repurposed, that, for example, features updated security algorithms
(crypto agility~\cite{ott_identifying_2019}) and cryptographic credentials to enable strong notions of
post-compromise security~\cite{cohn-gordon_post-compromise_2016}, and
generally following the principles laid out by Paverd et al.
in~\cite{pavert_sustainable_2019} can help us to sustain the secure and
safe operation of critical infrastructures far beyond of what is currently
feasible. Security and safety will still not be ecologically sustainable in
isolation but they can be tamed to serve communities to guarantee people's
wellbeing and to operate within the planetary boundaries. With ICT being a
pervasive industry sector like no other, one that has dramatically shaped
the way we work and interact socially, that also contributes tremendously
to the perpetuation of economic growth and exploitation, we must reject
notions of technological development that do not subject their methods and
outcomes to a reflection in the context of international climate- and
intersectional justice, and that do not promote equitable access to
technological advancement.

\paragraph{Future Directions.} This paper has mostly asked question to
guide research that seeks to reflect and embed sustained safety and
security in the life cycle of sustainable ICT products. Next steps in this
direction should be based on case studies -- products or research -- and
seek to develop concrete criteria to assess if and how just and sustained
notions of safety and security are implemented.




\begin{acks}
This paper is a follow-up of the author's keynote on \enquote{Sustainable
Security: What do we sustain, and for whom?} at the Workshop on
Sustainability in Security, Security for Sustainability at DATE
2022.\footnote{Workshop on Sustainability in Security, Security for
Sustainability at DATE 2022: \url{http://sussec22.alari.ch/}} I would like
to thank Birgit Penzenstadler, Christoph Becker, and Jay Chen for their
constructive feedback on this paper. I also thank everyone who contributed
to the discussion and development of the topic at DATE 2022 and also at the
SICT Summer Schools on Sustainable ICT in 2020 and 2021.\footnote{SICT
Summer Schools on Sustainable ICT:
\url{https://www.sictdoctoralschool.com/}} This research is partially
funded by the Research Fund KU Leuven, the Flemish Research Programme
Cybersecurity.

\end{acks}

\bibliographystyle{ACM-Reference-Format}
\bibliography{bibliography.bib,sustainable-ict.bib}


\end{document}